\documentclass{article}
\usepackage{amsmath,graphicx,mlspconf,amssymb}
\usepackage{bm}
\usepackage{hyperref}
\hypersetup{
        colorlinks = true,
        allcolors = blue,
        breaklinks = true,
        linktocpage=true,
}
\usepackage{multirow}
\usepackage{booktabs}
\usepackage{balance}
\title{Trimodal Glioma Representation Alignment via\\Volumetric Contrastive Learning}

\name{Denise Marini, Eleonora Grassucci, Danilo Comminiello}
\address{Dept. of Information Engineering, Electronics, and Telecommunications, Sapienza University of Rome, Italy}

\begin{document}
%\ninept

% \makeatletter
%     \let\@oldmaketitle\@maketitle% Store \@maketitle
%     \renewcommand{\@maketitle}{\@oldmaketitle% Update \@maketitle to insert...
%     \vspace{-0,5cm}
%     \includegraphics[width=0.95\textwidth]{Figures/MLSP2025 Palette Romanista.pdf}
    
%     \caption{Illustration of the modality gap in the medical imaging domain. Triangles represent embeddings extracted by the image encoder, while diamonds represent embeddings from the text encoder. Colors indicate shared semantic meaning. In standard CLIP-based training, the resulting latent space shows a significant modality gap (i.e., embeddings from different modalities with the same meaning remain far apart). Our method introduces additional loss functions designed to reduce this gap and to align cross-modal embeddings closely based purely on the semantic meaning.}
%     \label{fig:fig0}\bigskip}% ... an image
% \makeatother

\maketitle

\begin{abstract}
Glioma grading and survival prediction require the integration of heterogeneous information collected at different spatial and biological scales. Histopathology describes tissue morphology, mRNA expression captures molecular activity, and magnetic resonance imaging provides a non-invasive view of tumor extent and radiological heterogeneity. Existing glioma prognosis models often combine only two of these sources, while their alignment objectives remain mostly pairwise. This paper introduces GLORIA, a novel trimodal framework for GLioma Omics - Radiology - hIstopathology Alignment. GLORIA processes whole-slide image regions, gene-expression profiles, and 3D MRI volumes through modality-specific encoders, projects them into a shared latent space, and aligns them with a Gramian contrastive loss that measures the volume spanned by the three modality embeddings. The aligned representations are fused through a cross-modal gating module and optimized jointly for three-class glioma grading and overall survival prediction. We evaluate GLORIA on a matched TCGA-GBM/LGG and BraTS21 cohort, comprising 132 patients with all three modalities. On the shared trimodal test set, GLORIA improves over the bimodal WSI-mRNA baseline in all the metrics considered.
\end{abstract}
%

%%%%%%%%%%%% INTRODUCTION %%%%%%%%%%%
\section{Introduction}
\label{sec:intro}

Glioma grading and survival prediction are naturally multimodal problems as the disease is expressed at different spatial and biological scales. Histopathology remains the direct observation of tumor tissue architecture, cellular morphology, necrosis, microvascular proliferation, and other phenotypes used in grading. Molecular profiling is also essential: modern glioma classification has moved from morphology alone toward integrated histological, molecular definitions,
% with IDH status, 1p/19q codeletion, TP53, TERT, 
and genome-wide molecular patterns carrying strong diagnostic and prognostic information \cite{tcga2008gbm,tcga2015lgg,horbinski2022who}. MRI contributes another view by measuring tumor extent, enhancement, edema, mass effect, and spatial heterogeneity throughout the brain before tissue is sampled; radiogenomic studies further show that quantitative MRI features can be associated with molecular subtype and outcome \cite{rathore2021mrihistology,fan2024radiogenomics}. 
% These modalities therefore do not provide redundant inputs.
WSI, mRNA, and MRI describe morphology, molecular activity, and radiological structure, respectively, and each leaves part of the disease process unobserved.

This complementarity has motivated a growing body of bimodal and multimodal learning methods in glioma and cancer prognosis. Recent works show that joint histology-genomic models can learn prognostic representations that are not available to unimodal pathology or omics models alone \cite{chen2022pathomic,chen2022porpoise}. Concurrently, other works introduce co-attention between whole slide images (WSIs) patches and genomic features for survival prediction in gigapixel slides \cite{chen2021mcat}, and subsequent WSI-genomic models have continued to refine cross-modal interaction mechanisms. In parallel, MRI-histology and radiology-omics studies have reported that imaging features provide complementary prognostic information to tissue-derived features \cite{rathore2021mrihistology,li2023hmcat}.
% A systematic review of deep learning for glioma prognostication found that studies comparing multimodal and unimodal models generally favored multimodal integration, while also noting heterogeneity in datasets, modalities, and reporting \cite{alleman2023glioma}.
These results support multimodal representation learning as a modeling direction, but they also show that most existing work remains organized around pairs of modalities with conventional bimodal loss functions \cite{Oord2018RepresentationLW, Radford2021LearningTV}: WSI with omics, MRI with clinical or molecular variables, or MRI with histology \cite{alleman2023glioma}. This bimodal procedure upper-bounds the performance in downstream tasks, which instead may benefit from additional information brought by more modalities.

% SG-Fusion is the closest starting point for the present work because it targets glioma prognosis with a WSI--mRNA architecture: a Swin Transformer branch extracts histopathology features, a graph convolutional network encodes selected gene-expression profiles, and cross-modal fusion supports survival analysis and tumor grading \cite{fu2024sgfusion}. This design captures a meaningful genotype--phenotype relation, but it does not exploit MRI when matched volumes are available. Adding MRI is not a cosmetic extension of the input space. MRI is a volumetric signal with multiple acquisition sequences and spatial context at the organ scale, whereas WSI is a gigapixel tissue image and mRNA is a patient-level molecular vector. A trimodal model must therefore solve two problems simultaneously: each modality needs an encoder appropriate to its signal structure, and the resulting representations must be brought into a shared space without collapsing the distinct information carried by each source.
\begin{figure*}
    \centering
    \includegraphics[width=1\linewidth]{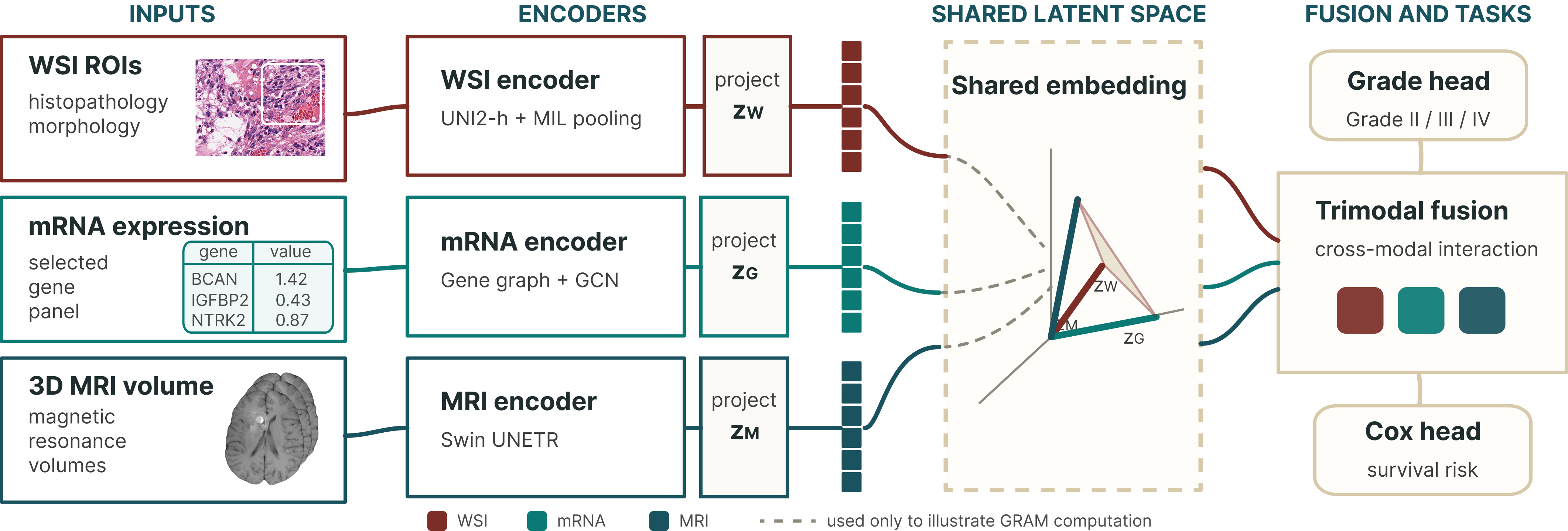}
    \caption{The GLORIA framework.}
    \label{fig:framework}
\end{figure*}

This paper proposes GLORIA: GLioma Omics - Radiology - hIstopatology Alignment, a novel framework going beyond bimodal approaches and incorporating three modalities (omics, radiology, and histopatology) altogether and leveraging their complementary information jointly. 
The method constructs a matched WSI-mRNA-MRI cohort by intersecting TCGA-GBM/LGG \cite{tcga2008gbm, tcga2008gbm} with BraTS21 \cite{baid2021brats}, processes each modality with a specialized encoder, and jointly aligns encoders embeddings through a volumetric contrastive loss function \cite{cicchetti2025gram}. In the framework, aligned embeddings span a smaller parallelotope in the shared latent space. This allows us to learn a patient representation in which histopathology, molecular activity, and radiological structure are coordinated and semantically meaningful. Such learned alignment among all the complementary modalities provide significant representations to the fusion module to perform downstream tasks, improving their performance. 
We evaluate the framework on two tasks: three-class glioma grading and overall survival prediction, demonstrating that jointly involving the three complementary modalities yields notable improvements in both tasks compared to the bimodal setting and previous methods.
\vspace{-0.8cm}
%%%%%%%%%%%% RELATED WORK %%%%%%%%%%%
\section{Related Work}
\label{sec:work}

Multimodal prognosis models have progressively combined tissue, molecular, radiological, and clinical information across oncology and glioma settings. Early histology-genomic approaches integrated pathology images with molecular or genomic measurements for outcome prediction \cite{mobadersany2018pnas}, while broader pan-cancer and pathology-omics models learned joint representations from clinical, molecular, and WSI-derived features or explicitly modeled histology-genomic interactions \cite{cheerla2019pancancer,chen2022pathomic,chen2022porpoise}. Attention-based methods further refined these interactions by linking WSI patch bags with genomic feature groups for survival prediction, including in brain tumor cohorts \cite{chen2021mcat,steyaert2023brain}. In parallel, glioma-focused radiology studies have combined multimodal MRI with survival targets, histological imaging, genomic information, and clinical variables through deep fusion, co-attention, orthogonal fusion, missing-modality learning, or transformer-based integration \cite{sun2019mri,rathore2021mrihistology,li2023hmcat,braman2021dof,cui2022missing,gomaa2024comprehensive}. Overall, systematic evidence suggests that multimodal deep learning often improves over unimodal modeling \cite{alleman2023glioma}, even in domains and applications that go beyond glioma prognosis \cite{Wang2022MedCLIPCL, Grassucci2025MLSP}.
Later, SG-Fusion targets glioma prognosis with a WSI-mRNA architecture \cite{FU2024102972} and, although the results are promising, the model is still limited to the bimodal setting. 

% \textbf{Multimodal Learning and Alignment Losses.} Starting from the cosine similarity-based CLIP losses \cite{Radford2021LearningTV} several multimodal models have been developed for
% two modalities like CLAP \cite{CLAP2022} or CLIP4Clip \cite{Luo2021CLIP4ClipAE}. Lately, the same loss has been extended to multiple modalities in ImageBind
% \cite{Girdhar2023ImageBindOE}, LanguageBind \cite{Zhu2023LanguageBindEV}, or VAST \cite{Chen2023VASTAV}. More recently, novel approaches have been proposed
% for multimodal learning to avoid the cosine similarity loss and rethinking multimodal alignment, namely GRAM \cite{cicchetti2025gram} that relies on the volume computation and Symile \cite{saporta2024contrasting}, based on total correlation. In the medical domain, the widespread method to align pairs of images and text is MedCLIP \cite{Wang2022MedCLIPCL}, which still relies on the conventional CLIP-based loss function.
\begin{table*}
    \caption{Patient-level evaluation on the shared trimodal test cohort ($n=29$), reported as mean$\pm$std over 7 seeds.
    % Best multimodal result per column in \textbf{bold}, best unimodal result \underline{underlined}.
    }
    \label{tab:main_results}
    \centering
    \setlength{\tabcolsep}{4pt}
    \resizebox{\textwidth}{!}{%
    \begin{tabular}{lccccccccc}
    \toprule
    Model & C-Index $\uparrow$ & Micro-F1 $\uparrow$ & Macro-F1 $\uparrow$ & Micro-AUC $\uparrow$ & Micro-AP $\uparrow$ & Composite $\uparrow$ & F1$_{\text{II}}$ $\uparrow$ & F1$_{\text{III}}$ $\uparrow$ & F1$_{\text{IV}}$ $\uparrow$ \\
    \midrule
    WSI unimodal & 0.813 $\pm$ 0.017 & 0.655 $\pm$ 0.040 & 0.702 $\pm$ 0.072 & 0.828 $\pm$ 0.016 & 0.662 $\pm$ 0.022 & 0.769 $\pm$ 0.007 & 0.633 $\pm$ 0.047 & 0.645 $\pm$ 0.048 & 0.829 $\pm$ 0.167 \\
    mRNA unimodal & \underline{0.891} $\pm$ 0.026 & \underline{0.729} $\pm$ 0.037 & \underline{0.803} $\pm$ 0.029 & \underline{0.897} $\pm$ 0.012 & \underline{0.804} $\pm$ 0.024 & \underline{0.844} $\pm$ 0.019 & \underline{0.678} $\pm$ 0.065 & \underline{0.730} $\pm$ 0.041 & \textbf{1.000} $\pm$ 0.000 \\
    MRI unimodal & 0.738 $\pm$ 0.061 & 0.424 $\pm$ 0.068 & 0.360 $\pm$ 0.091 & 0.592 $\pm$ 0.119 & 0.439 $\pm$ 0.081 & 0.615 $\pm$ 0.037 & 0.521 $\pm$ 0.092 & 0.288 $\pm$ 0.143 & 0.271 $\pm$ 0.212 \\
    \midrule
    SG-Fusion & 0.878 $\pm$ 0.022 & 0.675 $\pm$ 0.056 & 0.756 $\pm$ 0.045 & 0.873 $\pm$ 0.019 & 0.757 $\pm$ 0.028 & 0.816 $\pm$ 0.025 & 0.606 $\pm$ 0.138 & 0.662 $\pm$ 0.068 & \textbf{1.000} $\pm$ 0.000 \\
    Bimodal augm & \textbf{0.898} $\pm$ 0.022 & 0.690 $\pm$ 0.056 & 0.764 $\pm$ 0.050 & 0.860 $\pm$ 0.021 & 0.735 $\pm$ 0.040 & 0.828 $\pm$ 0.018 & 0.569 $\pm$ 0.124 & 0.724 $\pm$ 0.033 & \textbf{1.000} $\pm$ 0.000 \\
    GLORIA (ours) & 0.888 $\pm$ 0.033 & \textbf{0.749} $\pm$ 0.043 & \textbf{0.816} $\pm$ 0.033 & \textbf{0.894} $\pm$ 0.017 & \textbf{0.793} $\pm$ 0.036 & \textbf{0.848} $\pm$ 0.019 & \textbf{0.708} $\pm$ 0.076 & \textbf{0.741} $\pm$ 0.051 & \textbf{1.000} $\pm$ 0.000 \\
    \bottomrule
    \end{tabular}%
    }
\end{table*}

%%%%%%%%%%%% METHOD %%%%%%%%%%%
\section{Proposed Method}
\label{sec:method}
We propose GLORIA: GLioma Omics - Radiology - hIstopatology Alignment, a novel trimodal framework for glioma classification and survival prediction that integrates histopathology, transcriptomics and volumetric imaging into a shared representation space. The framework, shown in Fig.~\ref{fig:framework}, exploits the complementary information carried by the three modalities to improve predictive performance over common bimodal comparisons.

\subsection{Background}
Prior glioma fusion models such as SG-Fusion \cite{FU2024102972} combine whole-slide images (WSI) and mRNA expression in a bimodal framework with parallel encoders and a contrastive loss function.
Such a loss operates on a WSI-only triplet term: for each sample, an anchor $\mathbf{a}_i$ (the original ROI), a positive $\mathbf{p}_i$ (an augmented view of the same ROI), and a negative $\mathbf{n}_i$ (another sample in the batch) are encoded, with
\begin{equation}
\mathcal{L}_{\text{contr}} = \frac{1}{B}\sum_{i=1}^{B}\Big[\big(1-\langle \mathbf{a}_i,\mathbf{p}_i \rangle \big) + \mathrm{ReLU}\big(\langle \mathbf{a}_i,\mathbf{n}_i \rangle -m\big)^2\Big],
\label{eq:bimodal_contr}
\end{equation}
with margin $m=0.2$, $B$ the batch size, and $\langle\cdot,\cdot\rangle$ dot product, corresponding to the cosine similarity between two normalized vectors.
The total loss for the bimodal model (WSI and mRNA) further includes a contrastive and a PCA regularization term:
\begin{equation}
\mathcal{L}_{\text{bim}} = \mathcal{L}_{\text{grade}} + \mathcal{L}_{\text{hazard}} + \mathcal{L}_{\text{contr}} + \mathcal{L}_{\text{PCA}}.
\end{equation}
The proposed GLORIA framework extends SG-Fusion \cite{FU2024102972} along three axes: the addition of a third imaging modality, a dedicated volumetric encoder and a joint cross-modal alignment objective that treats the three modalities together. 

\subsection{Trimodal Integration}
While prior glioma fusion models combine whole-slide images (WSI) and mRNA expression, the radiological phenotype captured by MRI is complementary to both: it reflects macroscopic tumor structure that is not accessible from tissue sections or molecular profiles. We therefore add a third stream that processes 3D multi-parametric MRI alongside the histology and genomic branches, so that the fused representation draws on microscopic, molecular and macroscopic views of the same tumor.

% \subsection{Modality-specific encoders}
Each modality is encoded by a dedicated network suited its structure: a UNI2-h histopathology foundation model \cite{chen2024uni} with attention-based multiple-instance pooling \cite{ilse2018abmil} for WSI, a graph convolutional encoder over a gene-gene interaction graph for mRNA and the SwinViT encoder of a BraTS21-pretrained SwinUNETR \cite{hatamizadeh2022swinunetr} for MRI. The three encoders map their respective inputs into a common $d$-dimensional embedding space. Each embedding is first layer-normalized, $\tilde{\mathbf{z}}_k = \mathrm{LN}_k(\mathbf{z}_k)$ for $k\in\{\text{w},\text{m},\text{r}\}$, and projected through a modality-specific transform:
\begin{equation}
    \mathbf{h}_k=\tanh (W_k, \tilde{\mathbf{z}}_k) \in \mathbb{R}^d.
\end{equation}
Then, a cross-modal gate, conditioned on all three normalized embeddings $\tilde{\mathbf{z}}=[\tilde{\mathbf{z}}_{\text{w}};\tilde{\mathbf{z}}_{\text{m}};\tilde{\mathbf{z}}_{\text{r}}]$ produces an independent sigmoid score for every modality and every dimension, 
\begin{equation}
    \mathbf{g}_k=\sigma(V_k, \tilde{\mathbf{z}}) \in \mathbb{R}^d
    \qquad
    \boldsymbol{\alpha}_k=\frac{\mathbf{g}_k}{\sum_{k'\in\{\text{w},\text{m},\text{r}\} } \mathbf{g}_{k'}+\epsilon},
\end{equation}
\noindent where the division is element-wise so that the gates sum to one along the modality axis at each dimension. The fused representation is the per-feature convex combination, followed by a final layer norm:

\begin{equation}
    \mathbf{z}_{\text{fused}}=\mathrm{LN}(\sum_{k\in\{\text{w},\text{m},\text{r}\}} \boldsymbol{\alpha}_k \odot \mathbf{h}_k) \in \mathbb{R}^d,
\end{equation}

\noindent with $\odot$ the Hadamard product. Per-feature gating lets that each feature selects its own mixture. The fused embedding is passed to the grade-classification and hazard heads.

\subsection{Joint cross-modal alignment}
Rather than aligning modalities through a sum of pairwise cosine-similarity losses, which scales poorly with the number of modalities and treats each pair independently \cite{Chen2023VASTAV, cicchetti2025triangle}, we exploit the Gramian contrastive loss \cite{cicchetti2025gram}. This objective is designed to measure alignment as a geometric property of the joint embedding space spanned by all three modalities at once, aiming to capture higher-order relations that pairwise terms cannot express. This allows us to handle the three modalities jointly within a single objective rather than as separate pairwise constraints as common practices \cite{Girdhar2023ImageBindOE, Zhu2023LanguageBindEV}. Given the embeddings x, y, z $\in \mathbb{R^d}$ normalized to unitary norm, with mRNA as anchor, the similarity of such a triplet is the volume of the parallelotope they span, obtained from the determinant of their Gram matrix $\mathbf{G}$: 
\begin{equation}
\mathbf{G}(\mathbf{x},\mathbf{y},\mathbf{z})=
\begin{bmatrix}
\langle\mathbf{x},\mathbf{x}\rangle & \langle\mathbf{x},\mathbf{y}\rangle & \langle\mathbf{x},\mathbf{z}\rangle \\
\langle\mathbf{y},\mathbf{x}\rangle & \langle\mathbf{y},\mathbf{y}\rangle & \langle\mathbf{y},\mathbf{z}\rangle \\
\langle\mathbf{z},\mathbf{x}\rangle & \langle\mathbf{z},\mathbf{y}\rangle & \langle\mathbf{z},\mathbf{z}\rangle
\end{bmatrix}.
\end{equation}
\begin{figure*}[t]
    \centering
    \includegraphics[width=\textwidth]{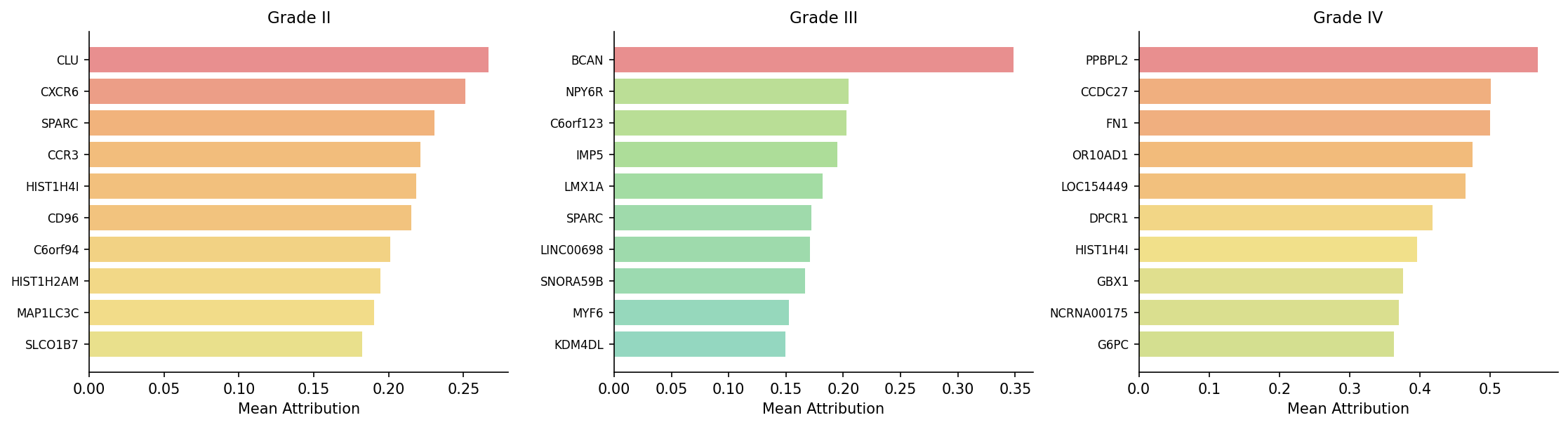}
    \caption{Top ten class-level gene signatures derived from mean attribution scores
    on the mRNA branch for each grade.}
    % (II, III, IV), the ten genes with the
    % highest mean attribution are shown.}
    \label{fig:gene_signatures}
\end{figure*}
\begin{figure}[t]
    \centering
    \includegraphics[width=1\columnwidth]{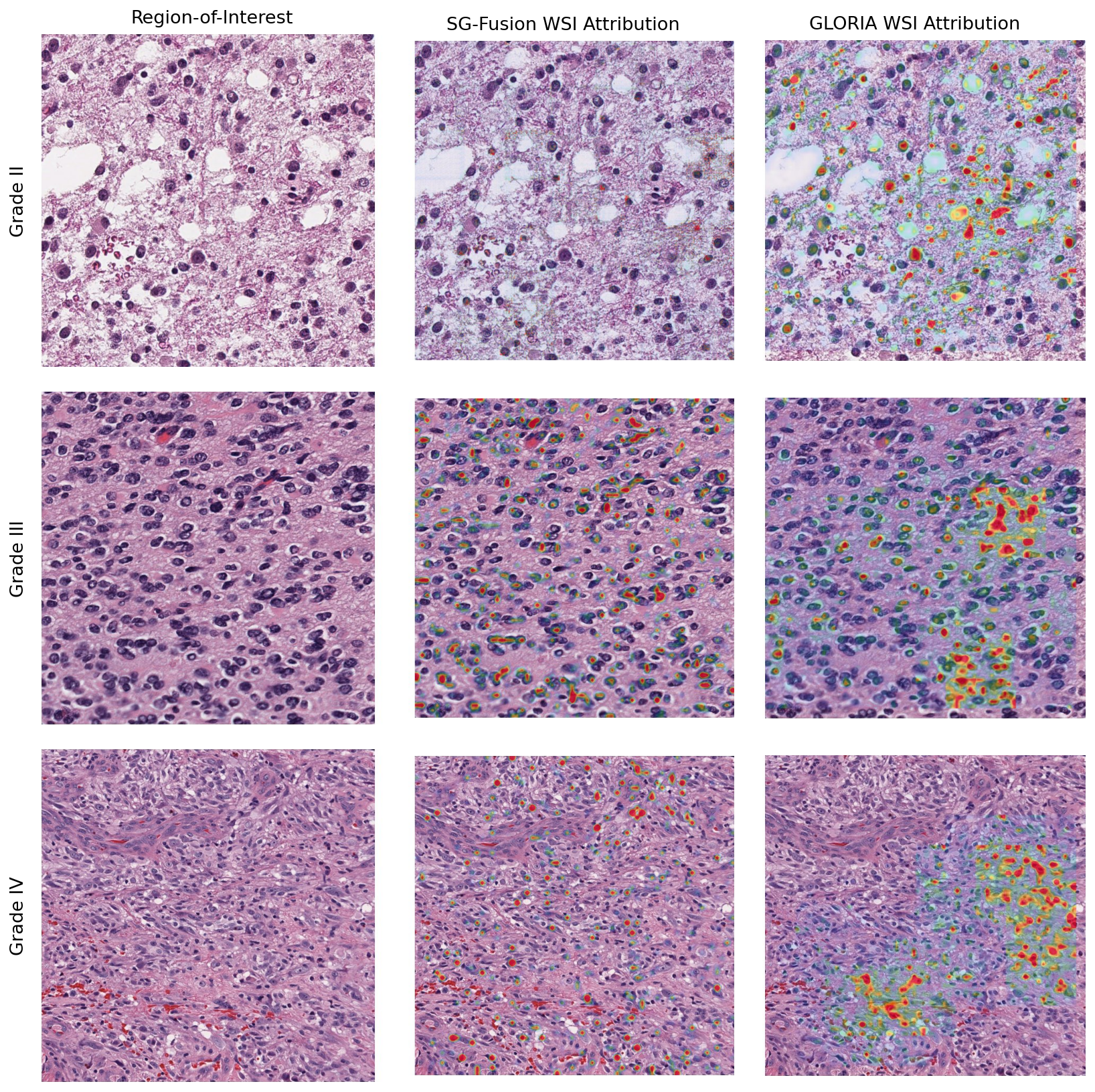}
    \caption{Qualitative comparison of WSI attribution between SG-Fusion and GLORIA across glioma grades. Rows correspond to grade II, III and IV.
GLORIA produces more spatially coherent attributions concentrated on
tumor-relevant regions, while SG-Fusion yields a sparse and diffuse, and thus less interpretable, response.}
    \label{fig:wsi_attributions}
\end{figure}
\noindent The volume is then obtained through the square root as:
\begin{equation}
\mathrm{Vol}(\mathbf{x},\mathbf{y},\mathbf{z})=\sqrt{\det \mathbf{G}(\mathbf{x},\mathbf{y},\mathbf{z})}.
\end{equation}
A smaller volume indicates a more aligned triplet. The volume replaces cosine similarity inside a symmetric InfoNCE objective \cite{Oord2018RepresentationLW}: for a batch of $B$ patients the volume of the anchor of sample $i$ with the second modality and third modality embeddings of sample $j$ defines the logit $\ell_{ij}=-\mathrm{Vol}(\cdot)/\tau$, trained in the data-to-anchor and anchor-to-data directions:
\begin{equation}
\mathcal{L}_{\text{D2A}} = -\frac{1}{B}\sum_{i=1}^{B}\log\frac{\exp(-\mathrm{Vol}(\mathbf{x}_i,\mathbf{y}_{i},\mathbf{z}_{i})/\tau)}{\sum_{j=1}^{B}\exp(-\mathrm{Vol}(\mathbf{x}_j,\mathbf{y}_{i},\mathbf{z}_{i})/\tau)},
\end{equation}
\begin{equation}
\mathcal{L}_{\text{A2D}} = -\frac{1}{B}\sum_{i=1}^{B}\log\frac{\exp(-\mathrm{Vol}(\mathbf{x}_i,\mathbf{y}_{i},\mathbf{z}_{i})/\tau)}{\sum_{j=1}^{B}\exp(-\mathrm{Vol}(\mathbf{x}_i,\mathbf{y}_{j},\mathbf{z}_{j})/\tau)},
\end{equation}
\begin{equation}
\mathcal{L}_{\text{GRAM}} = \tfrac{1}{2}\big(\mathcal{L}_{\text{D2A}} + \mathcal{L}_{\text{A2D}}\big),
\end{equation}
\noindent where $\mathbf{x}_i$, $\mathbf{y}_{i}$, $\mathbf{z}_{i}$ are the anchor (mRNA), WSI and MRI embeddings of patient $i$ and $\tau=0.07$ is the temperature. ROIs belonging to the same patient are masked out of the denominator to avoid treating them as false negatives.

\subsection{Training objective} The proposed model is trained jointly on two tasks: glioma grading, a three-class classification over Grade II/III/IV, and overall survival prediction via a Cox proportional-hazards objective \cite{FU2024102972}. This results in task losses $\mathcal{L}_{\text{grade}}$ and $\mathcal{L}_{\text{hazard}}$. The two tasks are balanced through learned homoscedastic uncertainty weighting: each task loss $\mathcal{L}_t$, with $t\in\{\text{grade},\text{hazard}\}$, is weighted as
\begin{equation}
\label{eq:loss_unc}
\mathcal{L}_t^{\text{w}} = \tfrac{1}{2}e^{-s_t}\mathcal{L}_t + \tfrac{1}{2}s_t, \qquad s_t = \log\sigma_t^2,
\end{equation}
where $s_t$ is a learnable parameter. We denote the combined supervised term as $\mathcal{L}_{\text{sup}} = \mathcal{L}_{\text{grade}}^{\text{w}} + \mathcal{L}_{\text{hazard}}^{\text{w}}$. 
%
% The total loss for the bimodal model (WSI and mRNA) further includes a contrastive and a PCA regularization term:
% \begin{equation}
% \mathcal{L}_{\text{total}} = \mathcal{L}_{\text{sup}} + \mathcal{L}_{\text{contr}} + \mathcal{L}_{\text{PCA}}.
% \end{equation}
% The contrastive loss operates on a WSI-only triplet term: for each sample, an anchor $a_i$ (the original ROI), a positive $p_i$ (an augmented view of the same ROI) and a negative $n_i$ (another sample in the batch) are encoded, with
% \begin{equation}
% \mathcal{L}_{\text{contr}} = \frac{1}{B}\sum_{i=1}^{B}\Big[\big(1-\cos(a_i,p_i)\big) + \mathrm{ReLU}\big(\cos(a_i,n_i)-m\big)^2\Big],
% \end{equation}
% with margin $m=0.2$ and $B$ the batch size.
In GLORIA, such losses are then supported with the joint contrastive loss:
\begin{equation}
\mathcal{L}_{\text{total}} = \mathcal{L}_{\text{sup}} + \lambda_{\text{GRAM}}\mathcal{L}_{\text{GRAM}},
\end{equation}
with $\lambda_{\text{GRAM}}=0.7$, keeping the same uncertainty-weighted supervised term $\mathcal{L}_{\text{sup}}$ of Eq.~\eqref{eq:loss_unc} for the tasks.
%
%%%%%%%%%%%% EXPERIMENTS %%%%%%%%%%%
\section{Experimental Results}
\label{sec:exp}

\textbf{Settings.} 
To assess the contribution of trimodal representation learning for glioma grading and survival prediction, we build a trimodal cohort by intersecting TCGA-GBM/LGG \cite{tcga2008gbm, tcga2015lgg}, for whole slide histopathology images and mRNA gene expression, with BraTS21 \cite{baid2021brats,bakas2017advancing,menze2015multimodal}, which provides the corresponding 3D MRI volumes. In this scenario, the cohort comprises 577 patients with paired WSI and mRNA, of which 132 also have available MRI, forming the trimodal subset. Each modality is processed as follows. \textit{WSI}: regions of interest (ROIs) of size 1024 are extracted from the diagnostic whole slide images. During training, a $512 \times 512$ window is randomly cropped from each ROI as spatial augmentation and to reduce the computational cost; at inference time, the full ROI is used. \textit{mRNA}: per patient, a pre-selected panel of 5724 genes is used, modeled as a graph in which genes are nodes and edges are given by a fixed gene-gene adjacency matrix \cite{FU2024102972}. \textit{MRI}: the four sequences (T1, T1ce, T2, FLAIR) are cropped to the tumor bounding box derived from the BraTS21 segmentation mask (16-voxel margin), resampled to $64^3$ via trilinear interpolation and z-score normalized per channel, resulting in a $4\times64^3$ volume. To process such heterogeneous data and obtain embeddings, we employ a dedicated encoder for each modality. For the WSI modality, we adopt UNI2-h \cite{chen2024uni}, a ViT-based histopathology foundation model. Each cropped patch is partitioned into $224 \times 224$ tiles via a sliding window (16 tiles for a $512^2$ training crop, 81 for the full ROI at inference), independently encoded, and aggregated into a single embedding through attention-based multiple-instance pooling \cite{ilse2018abmil}; formally, the WSI encoder is $e_W:\mathbb{R}^{224\times 224\times3}\rightarrow \mathbb{R}^d$. The mRNA modality is processed by the graph convolutional encoder inherited from \cite{FU2024102972}, in which the graph generated from the adjacency matrix is processed by two GCN layers, i.e., $e_G:\mathbb{R}^{5724} \rightarrow \mathbb{R}^d$. These two encoders are also used in the augmented bimodal model, together with the pairwise contrastive loss function defined in \eqref{eq:bimodal_contr}. For the MRI modality, we leverage the SwinViT encoder of a SwinUNETR model pretrained on BraTS21 \cite{hatamizadeh2022swinunetr}, discarding the decoder and appending a trainable projection head on the pooled multi-scale features, i.e., $e_M:\mathbb{R}^{4 \times 64^3} \rightarrow \mathbb{R}^d$. All three encoders project into a common latent space, whose dimensionality $d$ is set to 64. Both the augmented bimodal model and the proposed trimodal one are optimized with Adam. In the bimodal model, the pretrained WSI backbone blocks are fine-tuned with $lr=1\cdot10^{-5}$, while all remaining modules with $lr=1\cdot10^{-4}$. The trimodal model, initialized from bimodal weights, uses: $lr=5\cdot10^{-4}$ for pretrained WSI and mRNA encoders, $lr=1\cdot10^{-5}$ for the MRI backbone, and $lr=2\cdot10^{-3}$ for the new modules.
The comparison with the augmented bimodal model allows us to isolate the contribution of the third modality with the proposed joint contrastive loss in the GLORIA framework.

% The model is trained jointly on two tasks: glioma grading, a three-class classification over Grade II/III/IV, and overall survival prediction via a Cox proportional-hazards objective \cite{FU2024102972}.  The total loss for the bimodal model (WSI and mRNA) is: $\mathcal{L}_{total}=\mathcal{L}_{grade}+\mathcal{L}_{hazard}+\mathcal{L}_{contr}+\mathcal{L}_{PCA}$ , where the two tasks are balanced through learned uncertainty weighting $\frac{1}{2}e^{-s_t}\mathcal{L}_t+\frac{1}{2}s_t$ where $s_t=log\sigma^2_t $. The contrastive loss operates on WSI-only term in triplet form: for each sample, an anchor $a_i$ (the original ROI), a positive $p_i$ (an augmented view of the same ROI) and a negative $n_i$(another sample in the batch) are encoded. The loss is: $\mathcal{L}_{contr}=\frac{1}{B}\sum^B_{i=1}[(1-cos(a_i,p_i))+ReLU(cos(a_i,n_i)-m)^2]$ with margin $m=0.2$ and $B$ batch size. For the trimodal setup the total loss is defined in this way: $\mathcal{L}_{total}=\mathcal{L}_{grade}+\mathcal{L}_{hazard}+\lambda_{GRAM}\mathcal{L}_{GRAM}$, with $\lambda_{GRAM}=0.7$ and the same learned uncertainty weighting strategy of the bimodal case for grade and hazard losses. Cross-modal alignment is encouraged by the GRAM contrastive loss \cite{cicchetti2024gram}, applied to the encoder embeddings with the mRNA modality as anchor; same-patient regions of interest are masked out to avoid treating them as contrastive negatives.

\textbf{Metrics.} Models are evaluated at the patient level by aggregating the predictions over each patient's ROI. Grade classification is assessed by micro-averaged F1 and micro AUC, survival prediction by C-Index. To account for the strong class imbalance, the macro-averaged F1 and the per-class F1 are also reported. Performance is summarized by a composite metric, $0.5 \cdot C-Index + 0.3 \cdot F1 + 0.2 \cdot AUC$.

\textbf{Results.} Tab.~\ref{tab:main_results} reports the patient-level evaluation of SG-Fusion \cite{FU2024102972}, bimodal and trimodal models joint with the unimodal baselines on the shared cohort of 29 trimodal test patients, averaged over 7 seeds. With respect to SG-Fusion, GLORIA improves classification consistently across every aggregate metric 
% : micro-F1 increases by $+0.074$, macr-F1 by $0.060$, micro-AUC by $+0.021$ and micro-AP by $0.036$,
for a consistent composite gain, while survival ranking has a slight improvement. The per-class breakdown shows that the gain is concentrated on the two lower grades.
% , whose F1 rises by $+0.102$ (Grade II) and $+0.079$ (Grade III).
The bimodal model we augmented, by contrast, has small improvements on the metrics.
% C-Index ($+0.020$), micro-F1 ($+0.015$) and composite ($+0.012$).
Therefore, the uniform advantage over SG-Fusion emerges mostly once the volumetric MRI modality is added. Furthermore, adding the volumetric MRI modality consistently improves grade classification over the augmented bimodal model too.
% : micro-F1 increases by $+0.059$, macro-F1 by $+0.052$ and micro-AUC by $+0.034$. The improvement is largest and more robust for Grade II, whose F1 increases by $+0.139$:
The per-class analysis shows that MRI primarily resolves the systematic under-classification of Grade II at the II/III boundary, recovering cases that the bimodal model assigns to Grade III. This is a known and dominant source of error in glioma classification. Survival ranking, in contrast, is marginally lower for GLORIA (C-Index $-0.010$), indicating that the contribution of MRI concentrates on classification rather than prognosis. Fig.~\ref{fig:gene_signatures} reports the per-grade gene signatures, i.e. the top-10 genes ranked by mean attribution on the mRNA branch. The recovered signatures include known glioma-associated genes (CLU, SPARC, FN1), confirming that the genomic modality contributes biologically plausible evidence to the grade decision rather than spurious correlations. Fig.~\ref{fig:wsi_attributions} shows Grad-CAM attribution maps on the WSI encoder for one representative ROI per grade. Across all three grades, GLORIA concentrates its attribution on compact, nucleus-dense regions, with the strongest response on the hypercellular foci that drive the grade decision; in the Grade IV example, the map highlights the dense cellular core while attenuating the surrounding lower-density tissue. SG-Fusion, in contrast, produces a sparser and more diffuse response that is less clearly aligned with the discriminative regions. This qualitative difference is consistent with the quantitative gap in Tab.~\ref{tab:main_results}, and suggests that the multimodal alignment in GLORIA yields histological evidence that is both more localized and more interpretable.

%%%%%%%%%%%% CONCLUSION %%%%%%%%%%%
\section{Conclusion}
\label{sec:con}
This paper introduced GLORIA, a trimodal framework for glioma representation learning that jointly integrates histopathology, mRNA expression and 3D MRI. GLORIA aligns the three modality embeddings within a shared latent space before fusion. Experiments on TCGA-GBM/LGG and BraTS21 cohort show that adding MRI improves grade classification over the bimodal baseline, with gains in micro-F1, macro-F1, micro-AUC and the composite score, and with the largest improvement observed for Grade II cases. These results indicate that trimodal alignment is a promising direction for glioma modeling, while also motivating future validation on larger matched cohorts and further investigation of task-specific fusion strategies for prognosis.

\ninept
\bibliographystyle{IEEEtran}
\bibliography{biblio}
\balance

\end{document}